\documentclass[letterpaper]{article} 
\usepackage{aaai2026}  
\usepackage{times}  
\usepackage{helvet}  
\usepackage{courier}  
\usepackage[hyphens]{url}  
\usepackage{graphicx} 
\usepackage{natbib}  
\usepackage{caption} 
\usepackage{algorithm}
\usepackage{algorithmic}
\usepackage{tabularx}
\usepackage{booktabs}
\usepackage{amsmath}
\usepackage{adjustbox}
\newcommand{\answerYes}[1]{\textcolor{blue}{#1}} 
 
\newcommand{\answerNA}[1]{\textcolor{gray}{#1}} 
 
\usepackage{newfloat}
\usepackage{listings}
\usepackage{tikz}
\usetikzlibrary{shadows}
\usepackage{enumitem}
\usepackage[table]{xcolor}

\usepackage{newfloat}
\usepackage{listings}
\DeclareCaptionStyle{ruled}{labelfont=normalfont,labelsep=colon,strut=off} 
\floatstyle{ruled}
\newfloat{listing}{tb}{lst}{}
\floatname{listing}{Listing}
\makeatletter
\def\showauthors@on{T}
\makeatother
\title{Tied In on TikTok: Tie Strength and Emotional Dynamics in Algorithmic Communities}
\author{
  Charles Bickham\textsuperscript{\rm 1,2}, 
    Minh Duc Chu\textsuperscript{\rm 1,2}, 
    Arianna Yuan\textsuperscript{\rm 1},
    Valerie Lookingbill\textsuperscript{\rm 3},
    Ehsan Mohammadi\textsuperscript{\rm 3},
    Stuart Murray\textsuperscript{\rm 4},
  Kristina Lerman\textsuperscript{\rm 5}, 
  Emilio Ferrara\textsuperscript{\rm 1,2}
}
\affiliations{
  \textsuperscript{\rm 1}University of Southern California, Los Angeles, California, United States\\
  \textsuperscript{\rm 2}Information Sciences Institute, University of Southern California, Marina Del Rey, California, United States\\
  \textsuperscript{\rm 3}University of South Carolina, Columbia, South Carolina, United States\\
  \textsuperscript{\rm 4}University of California, Los Angeles, Los Angeles, California, United States\\
  \textsuperscript{\rm 5}Indiana University Bloomington, Bloomington, Indiana, United States\\
  cbickham@usc.edu, 
  dchu@isi.edu,
  yuana@usc.edu,
  lookingv@email.sc.edu,
  ehsa2@sc.edu,
  sbmurray@mednet.ucla.edu,
  lerman@isi.edu, emiliofe@usc.edu
}

\usepackage{bibentry}

\begin{document}

\maketitle

\begin{abstract}
Whether genuine communities can form on  algorithmically-driven short-form video platforms like TikTok remains an open question, given that user interactions are often brief, dispersed, and difficult to trace. Building on theories of tie strength and online community formation,  we examine whether eating disorder (ED) discourse on TikTok exhibits behavioral and emotional signatures of strong ties, including more frequent, reciprocal, and affectively intense  interactions. In this paper, we analyze 43,040 ED-related TikTok videos and over 560,000 comments, alongside a Non-ED comparison dataset. We find that at the user-pair level, greater interaction frequency is associated with increasingly positive emotional expression, a pattern that is amplified in ED-related conversations. This trend is also reflected linguistically, with pairs that interact more frequently exhibiting more of a positive tone. At the same time, how a relationship starts matters: pairs that begin with positive exchanges usually stay mostly positive as they continue interacting, while pairs that begin negatively may add some positive exchanges over time but rarely become mostly positive. To contextualize these dynamics, we classify ED videos into three content types (Pro-Recovery, Pro-ED, and ED Experiences) and find that each exhibits distinct emotional interaction patterns. These findings suggest that dense, emotionally structured relationships can emerge within ED discourse on TikTok. More broadly, our work provides one of the first empirical demonstrations of how community-like relational dynamics form and persist on algorithmically driven short-form video platforms.
\end{abstract}

\begin{links}
\link{Dataset}{https://github.com/cbickham3232/Tied-In-on-TikTok}
\end{links}

\section{Introduction}
Communities play a central role in social life, providing individuals with emotional support, shared identity, and access to information. Classic sociological theory conceptualizes communities as networks of \textit{strong social ties}---close, emotionally meaningful, and reciprocal relationships that bind individuals into cohesive groups~\cite{granovetter1973strength}. These ties sustain trust and belonging, forming the foundation of social support systems that help individuals navigate personal challenges and make sense of the world. 
Studies of large-scale communication networks provide empirical support for theories of tie strength: individuals who frequently and persistently interact with each other also share mutual contacts, signifying their embeddedness within dense communities~\cite{onnela2007structure}.

As social life has increasingly moved online, digital platforms have become critical spaces for community formation and support. Online communities can promote meaningful interaction, information sharing, and solidarity, particularly among vulnerable or stigmatized populations. For individuals facing mental health challenges~\cite{wang2017detecting,juarascio2010pro,wang2018social} or belonging to marginalized groups~\cite{clark2015black,jackson2018black,lubbers2022social}, social media can provide both a refuge and a risk. Eating disorder (ED) communities exemplify this duality: exposure to content glorifying eating disorders as a desirable aesthetic (i.e., pro-ED content) can disseminate harmful narratives, normalize disordered behaviors, and reinforce illness identities~\cite{bardone2007does,harper2008viewership,mulveen2006interpretative,custers2015urgent}; conversely, pro-recovery spaces allow users to share coping strategies, affirm progress, and offer mutual encouragement~\cite{au2022social,juarascio2010pro}.

Defining community on social media platforms, however, can be challenging. Social media platforms such as X (formerly known as Twitter) and Instagram rely on explicit social graphs, including follows and retweets, through which communities manifest as dense clusters of interpersonal connections~\cite{conover2011political,wang2018social,lerman2025safe}. These structures produce persistent, relationally anchored communities that reflect ideological, cultural, or emotional bonds. In contrast, TikTok’s social architecture is algorithmically curated rather than network-based. The platform’s ``For You Page'' (FYP) delivers personalized content through algorithmic recommendation rather than social ties. As a result, scholars argue that TikTok cultivates ``interest clusters''---ephemeral, algorithmically-linked collectives based on shared exposure to content rather than social relationships~\cite{lee2022algorithmic,gerbaudo2024tiktok,gombar2025algorithmic}. This leaves an open question whether genuine online communities can exist in an environment where users are connected primarily by algorithmic coincidence rather than intentional interaction.

This question is especially important in eating disorder related discourse, where repeated interactions can meaningfully influence emotional support, recovery, or risk. TikTok has become a major venue for ED content \cite{bickham2024hidden,herrick2021just,mccashin2023using,griffiths2024does}, especially among adolescents and emerging adults who face elevated risk for disordered eating behaviors \cite{pruccoli2022use,sauerwein2025scrolling,wokke2025impact,sjostrom2024helpful}. Algorithmic personalization further concentrates ED related content among vulnerable users \cite{griffiths2024does}, creating repeated opportunities for interaction that may shape both emotional tone and social dynamics over time. Whether these interactions remain superficial or develop into relationally meaningful exchanges has important implications for understanding social support, emotional contagion, and risk within algorithmically mediated spaces.

\textbf{Contributions of this work.}
In this paper, we provide an empirical examination of whether and how relational structure emerges on TikTok at the level of user dyads. Consistent with literature~\cite{cheng2024comprehensive}, we operationalize tie strength through repeated interaction between pairs of users and study how emotional expression evolves as interactions accumulate. Focusing on ED related discourse and a Non-ED dataset as baseline for comparison, we characterize how algorithmically mediated interaction gives rise to stable relational patterns in the absence of explicit social networks. Using a zero-shot prompting framework, we systematically categorize eating disorder–related content into distinct themes~\cite{lookingbill2023assessment}, including content promoting recovery vs  eating disorders.

Guided by this perspective, we address the following research questions:

\begin{itemize}
\item \textbf{RQ1:} How does increasing tie strength relate to emotional expression in user-user interactions on TikTok, and does this relationship differ in ED-related and Non-ED content?
\item \textbf{RQ2:} How is tie strength associated with positivity in ED-related user-pair interactions, and how does the initial interaction tone condition this relationship?
\item \textbf{RQ3:} Do recovery-oriented, pro-ED, and experience-sharing discussions exhibit different patterns of emotionally intense interaction as ties strengthen, and what do these differences reveal about the formation of strong ties within distinct ED-related communities?
\end{itemize}

Using 43,040 ED-related TikTok videos and over 560,000 comments, we show that emotional expression of interactions (comments) at the user-pair level becomes more positive as interaction frequency increases, with substantially more pronounced patterns in ED-related discourse than in Non-ED discourse. In other words, repeated exchanges are associated with a more positive tone. At the same time, early interactions play a defining role in these dynamics: relationships that begin positively tend to sustain positivity, whereas those that begin negatively may incorporate more positive exchanges over time but largely remain negative overall. These patterns are sensitive to the context of discourse, with 
different types of ED content exhibiting distinct emotional trajectories as ties strengthen.

Our findings challenge prevailing views of TikTok as a platform of purely transient engagement and demonstrate how community-like relational dynamics can emerge under algorithmic recommendation. More broadly, this work highlights how online communities continue to fulfill enduring social functions of support, identity, and belonging, even as the mechanisms organizing interaction shift from social networks to algorithmic curation.

\section{Related Works}
\paragraph{Social Ties \& Online Communities}
Prior work in sociology and network science has examined how communities emerge through strong social ties, typically defined by frequent and reciprocal interaction and structural embeddedness within networks~\cite{granovetter1973strength,onnela2007structure}. These concepts have been operationalized across a range of offline and online settings to study cohesion, information flow, and social support.

On platforms like X and Instagram, online communities are commonly identified through explicit social graphs, such as follow relationships or retweet networks. In these settings, dense clusters of interaction often correspond to shared interests, identities, or ideological alignment. For example, political communication on Twitter has been shown to organize into polarized echo chambers~\cite{conover2011political}. Similar network-based approaches have been applied to eating disorder related discourse, where retweet networks reveal tightly connected subcommunities, including both pro-ED groups that circulate potentially harmful content and pro-recovery groups oriented toward support and treatment~\cite{wang2018social,lerman2025safe}. In contrast, TikTok operates with limited reliance on explicit social connections, instead emphasizing algorithmic content recommendation. Prior research suggests that this form of personalization reshapes patterns of exposure and identity expression~\cite{lee2022algorithmic}. As a result, TikTok communities are often described as “interest clusters,” formed through shared algorithmic exposure and trend participation rather than persistent interpersonal ties~\cite{gerbaudo2024tiktok,gombar2025algorithmic}. 

\paragraph{Social Media \& Emotions}
Emotions are a central part of both daily interactions and engagement on social media. Research shows that people frequently mirror the emotions they encounter online \cite{goldenberg2020digital}. Emotional states such as happiness \cite{fowler2008dynamic}, loneliness \cite{cacioppo2009alone}, and depression \cite{rosenquist2011social, de2013predicting} have been shown to spread across social networks, shaping how individuals express themselves. Repeated exposure to a particular emotional tone increases the likelihood of responding in a similar way \cite{ferrara2015measuring}, and even something as simple as viewing a smiling selfie can raise viewers’ smile intensity by 15\% \cite{sasaki2021investigating}.

These dynamics are also evident in specific online communities. For example, participation in depression-related forums can intensify emotional experiences, where positive interactions promote hope while negative ones strengthen feelings of hopelessness and fear \cite{tang2021emotional}. Exposure to toxic content similarly provokes strong affective reactions, with more severe toxic replies linked to greater expressions of anger and sadness \cite{aleksandric2024users}. Emotional expression also influences patterns of engagement on social platforms. In cancer-related Twitter discussions, joy, sadness, and hope tended to receive more likes, while joy and anger were more likely to be amplified through retweets \cite{wang2020fear}.

\paragraph{Social Media \& Mental Health Communities} Prior research has examined how social media enables both supportive and harmful community dynamics around mental health. Frequent and sustained engagement has been linked to stronger online participation \cite{chhikara2024letstalk} and social connectedness among individuals with mental illness \cite{brusilovskiy2016social, gowen2012young}. In the context of EDs, scholars have analyzed how online communities form, interact, and evolve across platforms. Early work on X and Instagram revealed that pro-eating disorder networks display strong homophily and insular communication patterns, with recovery-oriented users forming smaller, less connected subgroups \cite{wang2017detecting, wang2018social}. Other studies conceptualize the growth of pro-ana discourse as an online radicalization process that reinforces harmful norms through repeated exposure \cite{lerman2023radicalized}. On Instagram, pro-ana and thinspiration content has become increasingly mainstreamed, embedding ideals of self-discipline and restriction within broader youth culture \cite{ging2018written}, while recovery spaces offer peer validation and representation but face risks of unmoderated harmful content \cite{au2022social}. Network analyses of ED discourse also highlight how influential users amplify unhealthy behaviors and emerging trends such as ``\#meanspo,'' reflecting the shifting language of these communities \cite{nova2022cultivating}. Finally, large-scale computational studies show that mental illness severity in pro-ED posts has increased over time, underscoring the urgency of understanding how community structures and repeated engagement shape online mental health discourse \cite{chancellor2016quantifying}.

\section{Methodology}

\subsection{Data}
For this study, we utilize our publicly-available TikTok eating disorder dataset~\cite{bickham2025edtokdataseteatingdisorder}. This collection contains 43,040 TikTok videos accompanied by 388,332 top-level comments and 188,379 replies, spanning the period from January 2019 to June 2024. To provide a comparison group, a Non-ED dataset was collected using the same keywords employed in collecting the EDTok corpus but applying a NOT operator to exclude eating disorder-related terms. The resulting Non-ED dataset consists of 37,530 videos, 203,709 original comments, and 102,902 replies. 

Data for the Non-ED dataset were retrieved via the TikTok Research API. Each video record includes a unique identifier, while each comment entry contains its own ID, the text content, the date of the comment and, where relevant, a reply ID. This structure makes it possible to reconstruct conversation threads and examine patterns of interaction.

In addition, we incorporate an independently annotated dataset of 200 TikTok videos from \cite{lookingbill2023assessment}, which we use as ground truth for evaluation. Each entry in this dataset contained the original mp4 video file, the unique video ID, and human-provided annotations labeling the video into one of four eating disorder–related content types. This dataset provides an external benchmark that allows us to evaluate classification performance and to calibrate our zero-shot prompt before scaling to the full dataset.

\subsection{Measuring Emotions}
To analyze emotional expression in the comments, we employed two emotion classification models: Demultiplexer (Demux) \cite{chochlakis2023leveraging} and RoBERTa-GoEmotions.\footnote{\url{https://huggingface.co/SamLowe/roberta-base-go_emotions}}
 Demux is a multi-label emotion recognition model derived from SpanEmo \cite{alhuzali2021spanemo}. It encodes two inputs—the emotion categories and the text sample—to generate contextual embeddings that yield probability scores across emotions. In contrast, RoBERTa-GoEmotions is a transformer-based classifier trained on the GoEmotions dataset \cite{demszky2020goemotions}, which contains 28 emotion classes. For comparability, we aligned its predicted categories with those of Demux and grouped each emotion by polarity (positive, neutral, or negative) to ensure consistent aggregation across models (see Table~\ref{tab:emotion-mapping} in the Appendix).

Both models were applied to the full set of comments, targeting emotions such as anger, anticipation, disgust, fear, joy, love, optimism, pessimism, sadness, surprise, trust, and none. To assess accuracy, we manually annotated a random sample of 500 comments. On this benchmark, Demux achieved F1 scores of 92.26\% for comments, while RoBERTa-GoEmotions performed considerably worse with a score of 33.62\% (see Table \ref{table:labelresults}). Agreement between the two models was also limited, with Cohen’s Kappa of 0.21 for comments and overall agreement rates of 27.80\%. Given its substantially stronger alignment with human annotations, Demux was selected as the primary model for emotion detection in this study.

\subsection{Linguistic Cues}
We applied the 2022 release of the Linguistic Inquiry and Word Count (LIWC) tool \cite{boyd2022development} to examine psychological and linguistic characteristics of the comments. LIWC is a widely used software package for text analysis that identifies the proportion of words in a document that fall into theoretically and empirically validated categories. These categories span both structural aspects of language (e.g., pronouns, articles, and function words) as well as psychological dimensions such as affective expression, cognitive processes, and social interaction. Access to LIWC requires a license, which we obtained for this research.

Our analysis concentrated on the Positive Tone and Negative Tone categories, which capture sentiment rather than discrete emotions. These dimensions include not only words directly linked to emotional states (e.g., happy, joy, sad, angry) but also terms associated with emotional contexts (e.g., birthday, beautiful, kill, funeral). For example, in the transcript ``I am very happy,'' one out of four words is coded as Positive Tone, resulting in a score of 25. LIWC was selected for this task because it provides a continuous lexical measure of sentiment that can be aggregated across multiple interactions, a property that has been shown to be effective for estimating aggregate sentiment in large collections of short social media texts. This property is particularly useful for relationship-level analysis, where tone must be summarized across many short comments exchanged between the same pair of users. Across our dataset, the mean Positive Tone score was 11.36 and the mean Negative Tone score was 2.01, both of which exceed the averages reported in prior Twitter-based analyses (6.05 and 1.85, respectively) \cite{boyd2022development}. These differences provide a comparative benchmark for interpreting the prominence of sentiment expression in TikTok eating disorder-related content.

\begin{table}[t]
\centering
\scriptsize
\resizebox{\columnwidth}{!}{%
\begin{tabular}{lccc ccc}
\toprule
 & \multicolumn{3}{c}{\textbf{Demux}} & \multicolumn{3}{c}{\textbf{RoBERTa-Go}} \\
\cmidrule(lr){2-4} \cmidrule(lr){5-7}
 & \textbf{P} & \textbf{R} & \textbf{F1} & \textbf{P} & \textbf{R} & \textbf{F1} \\
\midrule
Results & 92.77 & 92.20 & 92.26 & 47.33 & 27.59 & 33.62 \\
\bottomrule
\end{tabular}%
}
\caption{Precision, recall, and F1-score for Demux and RoBERTa-GoEmotions on the annotated sample of 500 comments.}
\label{table:labelresults}
\end{table}

\subsection{Content Classification}
\cite{lookingbill2023assessment} developed a taxonomy to categorize eating disorder–related content on TikTok, distinguishing harmful material that promotes eating disorders (pro-ED), videos sharing personal experiences with eating disorders, and content emphasizing social support and recovery (pro-recovery). This taxonomy enables a systematic assessment of eating disorder–related discourse on the platform.

The taxonomy was developed through thematic analysis of 200 TikTok videos related to eating disorders, which the authors coded to identify recurring patterns. From this process, they distilled 23 subthemes into four overarching categories: 
\begin{itemize}
    \item \textbf{Pro-ED Content}: Encouraging eating disorders by promoting them as a lifestyle choice, seeking support to maintain disordered behaviors, or urging others to adopt them.
    \item \textbf{Pro Recovery Content}: Sharing recovery stories by offering personal experiences, advice, encouragement, and challenges faced during the recovery process.
    \item \textbf{ED Experiences}: Sharing personal experiences with eating disorders to raise awareness or educate others, including how they developed and the physical and emotional symptoms they face.
    \item \textbf{Social Support}: Creators give, seek, or share experiences of receiving social support both online and offline.
\end{itemize}

We first applied Google Gemini\footnote{\url{https://gemini.google.com/}\label{note1}}
 in a zero-shot setting (see Figure \ref{fig:gemini_prompt} in the Appendix) to the existing annotated dataset of 200 videos from \cite{lookingbill2023assessment}. Classification performance on this dataset is reported in Table \ref{table:GroundTruthResults}. Based on this evaluation, we applied the same zero-shot prompt to label the full ED dataset. To further validate labeling performance on the ED data, two authors independently constructed a ground truth set of 200 videos. Classification results on this manually annotated subset are shown in Table \ref{table:EdtokResults}. Given the relatively low classification performance across both evaluation datasets, the Social Support category was excluded from subsequent analyses.

\begin{table}[t]
\centering
\scriptsize
\resizebox{\columnwidth}{!}{%
\begin{tabular}{lcccc}
\toprule
\textbf{Category} & \textbf{Accuracy} & \textbf{Precision} & \textbf{Recall} & \textbf{F1} \\
\midrule
Recovery Content & 78.26\% & 48.65\% & 78.26\% & 60.00\% \\
Pro-ED Content   & 82.69\% & 93.48\% & 82.69\% & 87.76\% \\
ED Experiences   & 63.22\% & 74.32\% & 63.22\% & 68.32\% \\
Social Support   & 15.38\% & 50.00\% & 15.38\% & 23.53\% \\
\bottomrule
\end{tabular}%
}
\caption{Ground truth classification results.}
\label{table:GroundTruthResults}
\end{table}

\begin{table}[t]
\centering
\scriptsize
\resizebox{\columnwidth}{!}{%
\begin{tabular}{lccccr}
\toprule
\textbf{Category}  & \textbf{Precision} & \textbf{Recall} & \textbf{F1-Score} & \textbf{\# Videos} \\
\midrule
Recovery Content  & 80.18\% & 95.70\% & 87.25\% & 22,425 \\
Pro-ED Content    & 66.67\% & 100.00\% & 80.00\% & 1,247  \\
ED Experiences    & 94.20\% & 78.31\% & 85.53\% & 11,987 \\
Social Support   & 75.00\% & 37.50\% & 50.00\% & 2,472  \\
\bottomrule
\end{tabular}%
}
\caption{Classification results on the EDTok dataset.}
\label{table:EdtokResults}
\end{table}

\section{Results}

\subsection{RQ1: Is Tie Strength Related to Stronger Emotional Expression in TikTok Interactions?}

We begin by operationalizing \textit{tie strength} at the level of user dyads as the total number of comments exchanged between two users across all videos in the dataset. This measure captures persistence of  interaction. The distribution of user–user exchanges in both datasets is highly skewed, with most pairs interacting only once but a small subset engaging in many repeated interactions (see Figure \ref{fig:ccdf} in the Appendix). Notably, the ED dataset exhibits a heavier tail than the Non-ED dataset, indicating a greater concentration of dyads with sustained, repeated interactions, with some user pairs interacting hundreds of times.  

To examine how emotional expression varies across these levels of relational intensity, we group user pairs into discrete interaction bins corresponding to increasing tie strength: 1, 2–5, 6–10, 11–19, and 20 or more exchanged comments. For each bin, we compute the aggregate emotional tone across all comments exchanged within those relationships.

\begin{figure*}[t]
    \includegraphics[width=\linewidth]{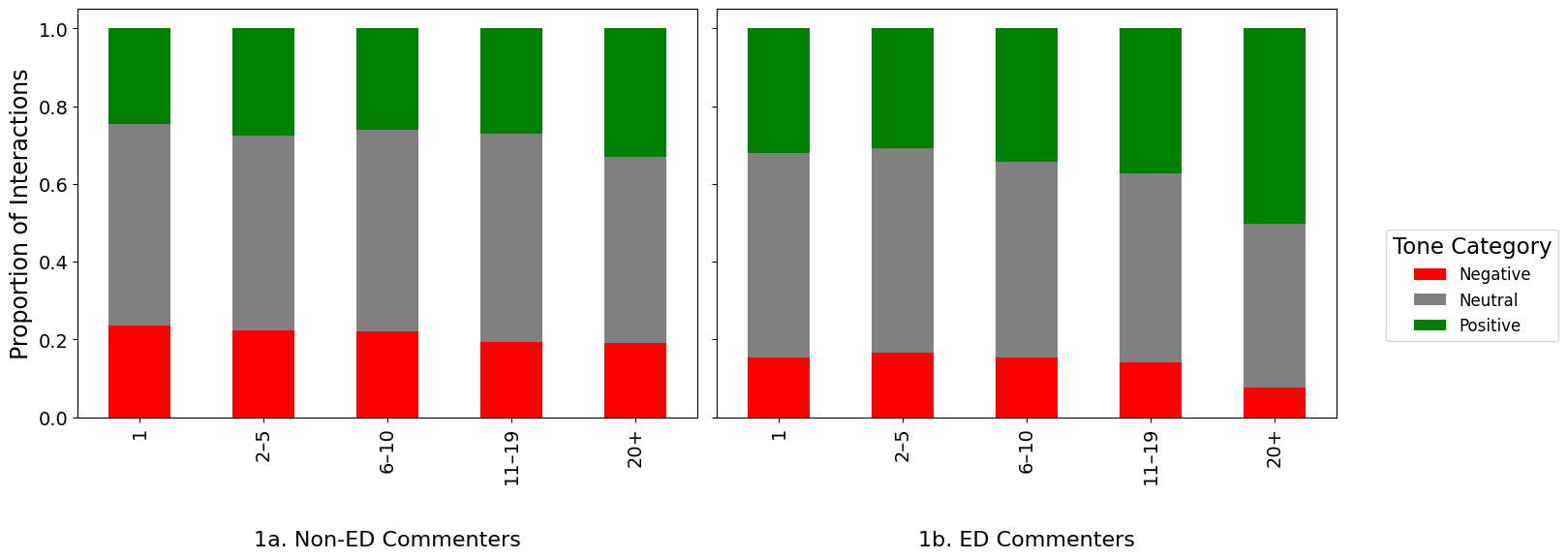}
    \caption{Distribution of emotional tone across interaction bins for Non-ED (1a) and ED (1b) commenters. Each bar represents the proportion of negative (red), neutral (gray), and positive (green) comments within a given interaction frequency range. Compared to the Non-ED baseline, ED commenters show a sharper decline in negativity and a stronger amplification of positivity as interaction frequency increases, particularly in the 20+ bin.}
    \label{fig:distrofemotions}
\end{figure*}

\begin{figure*}[t]
    \includegraphics[width=\linewidth]{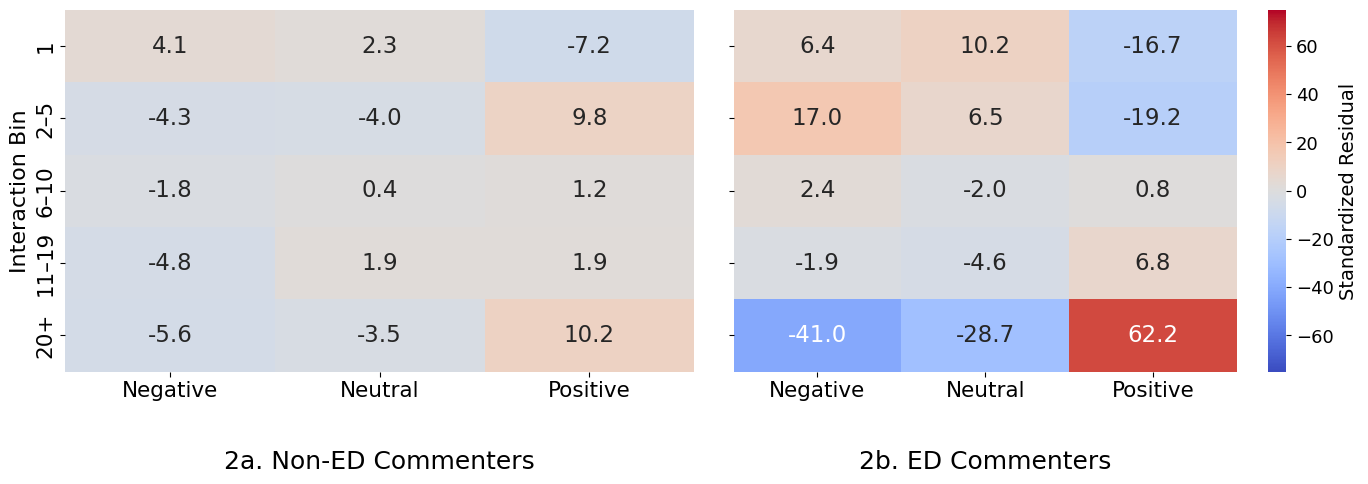}
    \caption{Standardized residuals from Chi-Square Tests of Independence for Non-ED commenters (2a) and ED commenters (2b). Positive values (red) indicate overrepresentation of tone categories relative to expectation, while negative values (blue) indicate underrepresentation. ED commenters show a stronger amplification of positivity and sharper underrepresentation of negativity at higher interaction bins (20+), compared to the more moderate shifts observed in Non-ED commenters.}
    \label{fig:chisquareNonED}
\end{figure*}

We analyze the Non-ED dataset to establish a baseline pattern. As shown in Figure \ref{fig:distrofemotions}a,  user-pairs that interact more frequently exhibit a higher share of positive emotional interactions and a corresponding decline in negative interactions. This trend indicates  that repeated interactions between the same users are increasingly characterized by positivity.
This association is supported by a Chi Square Test of Independence between interaction bins and emotional tone categories. Examination of the standardized residuals (Figure \ref{fig:chisquareNonED}a) shows that user-pairs with more than one interaction are overrepresented in the positive tone category, with the strongest effect observed in the 20+ bin (residual = 10.2, $p < 0.05$). These same user-pairs are underrepresented in the negative tone category (residual = -5.6, $p < 0.05$), indicating that increased interaction frequency is associated with a shift away from negative expression in Non ED discourse.

Next, we apply the same analysis to the ED dataset. Figure \ref{fig:distrofemotions}b shows a similar pattern, but with substantially stronger trend. User-pairs in ED discourse display a sharper increase in positive emotional tone and a more pronounced decline in negative tone with interaction frequency. The Chi Square Test of Independence further confirms this pattern. Standardized residuals for ED related interactions (Figure \ref{fig:chisquareNonED}b) reveal strong overrepresentation of positive tone among relationship pairs in the 20+ interaction bin (residual = 62.2, $p < 0.05$), alongside substantial underrepresentation of negative tone (residual = -41.0, $p < 0.05$). 

\textbf{Summary.}
Across both datasets, stronger tie-strength is associated with systematically more positive and less negative emotional expression at the relationship level. This association is present in Non-ED discourse but is significantly amplified in ED related interactions, particularly among frequently interacting user-pairs.

\subsection{RQ2: Linguistic Tone and Initial Interaction Effects in ED Relationships}
We next examine how the emotional tone and interaction trajectories vary across user pair relationships for ED TikTok content. To quantify emotional tone at the relationship level, we use LIWC-derived sentiment scores rather than the Demux classifier, since LIWC provides a continuous lexical signal that can be aggregated across many short interactions between the same pair of users. We compute a \textit{tone score} for each user pair $(i,j)$. Let $\mathcal{C}_{ij}$ be the set of all comments exchanged between $i$ and $j$ in either direction (i$\rightarrow$j or j$\rightarrow$i), and let $N_{ij} = |\mathcal{C}_{ij}|$. For each comment $c \in \mathcal{C}_{ij}$, let $p_c$ and $n_c$ denote the positive-tone and negative-tone values from LIWC. We define:

\[
\bar{p}_{ij} = \frac{1}{N_{ij}} \sum_{c \in \mathcal{C}_{ij}} p_c, \quad
\bar{n}_{ij} = \frac{1}{N_{ij}} \sum_{c \in \mathcal{C}_{ij}} n_c
\]

\[
\text{ToneScore}_{ij} = \bar{p}_{ij} - \bar{n}_{ij}
= \frac{1}{N_{ij}} \sum_{c \in \mathcal{C}_{ij}} \big(p_c - n_c\big)
\]

\begin{figure}[t]
    \includegraphics[width=\linewidth]{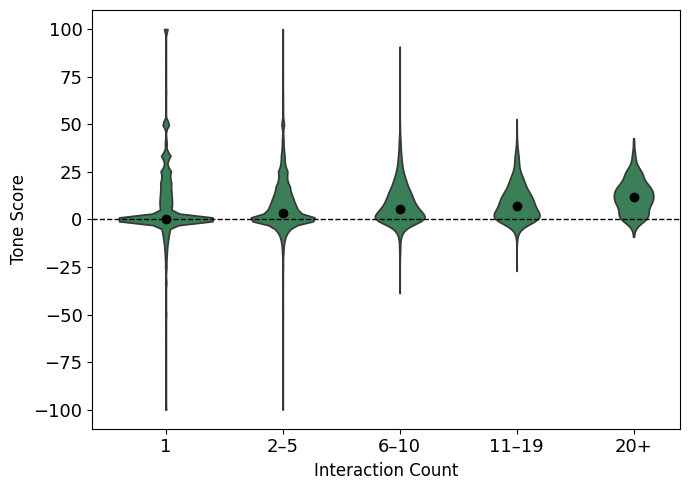}
    \caption{Tone score distributions across user pair relationships by interaction bin for ED TikTok content. Violin plots show the distribution of relationship-level tone scores within each bin, with black dots marking medians and the dashed line indicating neutrality. Distributions shift toward higher values as interaction frequency increases, reflecting more positive and more stable emotional expression.}
    \label{fig:tonescore}
\end{figure}

A positive score indicates that the pair’s exchanges are, on average, more positive than negative, while a negative score indicates the opposite. This formulation takes into account \textit{all} comments between each user pair, ensuring that interaction frequency is directly reflected in the score. As shown in Figure \ref{fig:tonescore}, tone scores generally shift upward and display less variance as the number of interactions increases. Importantly, once pairs surpass three reciprocal interactions (see Figure \ref{fig:threetonescore} in the Appendix), their tone scores are consistently positive. This suggests that repeated and reciprocated engagement between the same users reinforces positivity at the linguistic level.

To further contextualize these patterns, we examine how interaction patterns differ depending on whether a relationship begins with a positive or negative exchange in ED-related content. We restrict this analysis to user pairs with at least two interactions and classify each relationship as either positive-start or negative-start based on the tone of its initial exchange. For each group, we analyze interaction composition across increasing levels of tie strength.

\begin{figure}[t]
    \includegraphics[width=\linewidth]{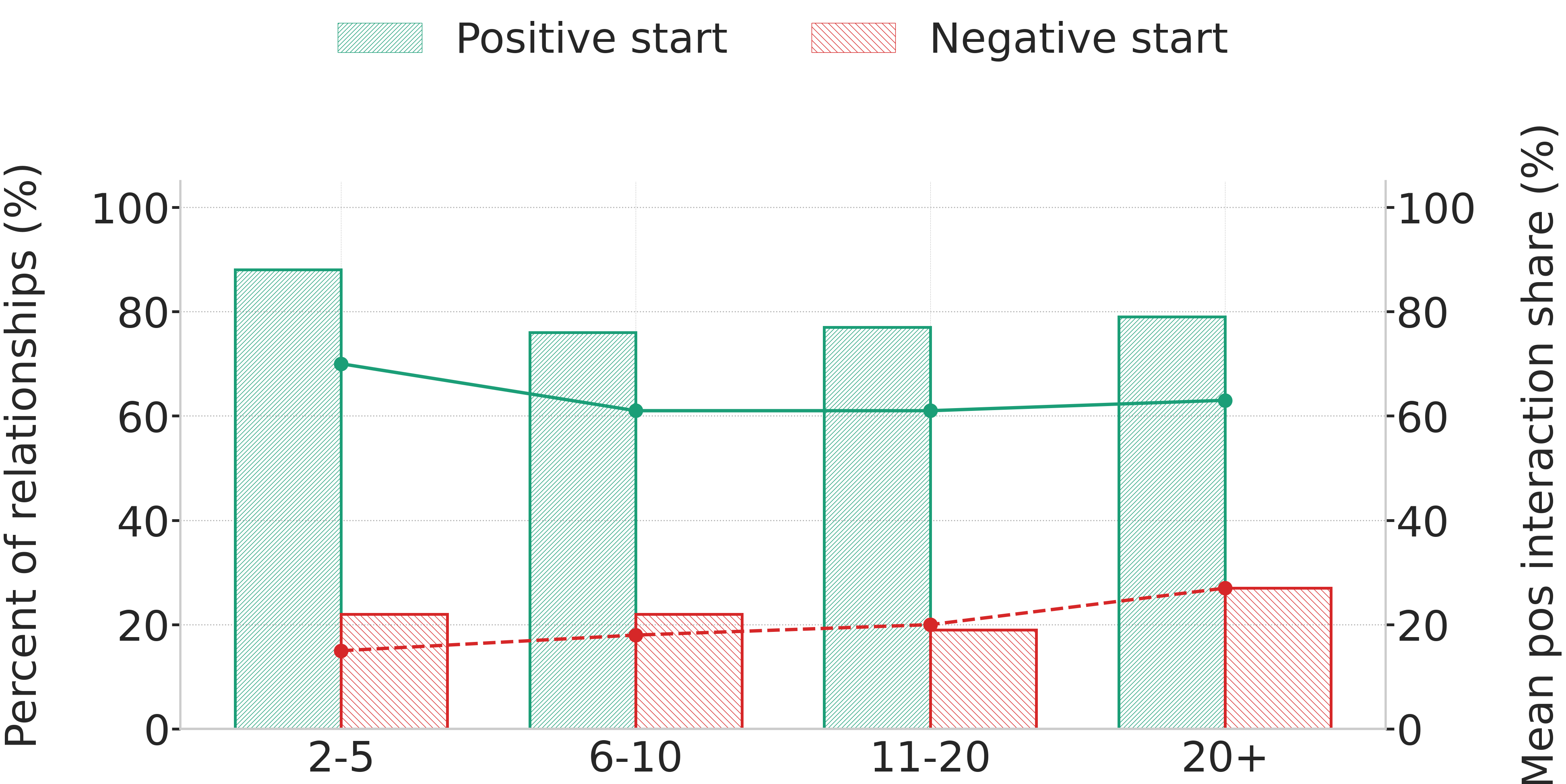}
    \caption{Relationship-level interaction patterns by initial emotional tone. Bars (left y-axis) show the percentage of relationships that are mostly positive at different interaction levels, while lines (right y-axis) show the average share of positive interactions.}
    \label{fig:start}
\end{figure}

For negative-start relationships relatively few pairs transition to positive majority, defined as cases where more than half of all interactions between two users are positive in tone. Across interaction bins, only a minority of negative-start relationships reach this threshold, even among pairs with high interaction frequency (see Figure \ref{fig:start}). This indicates that while negative-start relationships may incorporate positive exchanges over time, these interactions often coexist with sustained negativity rather than replacing it. In contrast, positive-start relationships are far more likely to maintain a positive majority as interactions accumulate. Even at lower interaction frequencies, most positive-start relationships remain above the majority-positive threshold, reflecting a tendency for early positivity to persist as the dominant emotional mode of interaction.

\textbf{Summary.}
Across user pair relationships, repeated interaction is associated with increasingly positive tone, indicating that tie strengthening reinforces affective alignment over time. However, early emotional conditions matter: relationships that begin positively tend to sustain positivity as interactions accumulate, while those that begin negatively rarely transition to a predominantly positive state despite incorporating some positive exchanges.

\subsection{RQ3: Differences in Relational Emotional Dynamics Across ED Discussion Spaces}
Lastly, we examine whether the emotional expression of frequent interactions differs across distinct ED-related discussion spaces. Following \cite{lookingbill2023assessment}, we categorize ED TikTok videos into three content types: Pro-Recovery, Pro-ED, and ED Experiences. For each category, we analyze how emotional expression evolves within \textit{user–user relationships} as tie strength increases. Figure \ref{fig:distrforcontent} in the Appendix shows the distribution of interaction frequencies across content types, highlighting substantial variation in the prevalence and depth of repeated dyadic interactions.

\subsubsection{Pro-Recovery Content}
Comments on Pro-Recovery videos often express affirmation, pride, reassurance, and mutual recognition of progress, including references to treatment milestones, fear food challenges, and personal recovery experiences. As shown in Figure \ref{fig:commchi}a, Pro-Recovery content exhibits a strong association between tie strength and emotional tone. Among user pairs with more than 20 interactions, positive emotion is substantially overrepresented (residual = +25.4, $p < 0.05$), while negative emotion is strongly underrepresented (residual = -25.0, $p < 0.05$). This emotional pattern shows that as relationships deepen in Pro-Recovery spaces, they also become stronger emotionally, consistent with formation of  strong social ties.

\subsubsection{Pro-ED}
In contrast, interactions around Pro-ED content exhibit a markedly different emotional profile. Comments between recurring user pairs primarily involve transactional or informational exchanges, such as calorie counts, weight references, or body comparisons—rather than affective engagement. The overall tone is less emotional compared to other types of ED content. As shown in Figure \ref{fig:commchi}b, standardized residuals remain close to zero across all interaction bins. Even among the most frequently interacting pairs, emotional expression does not systematically intensify in either a positive or negative direction (negative residual = -2.3, $p < 0.05$). These patterns suggest that repeated interaction in Pro-ED spaces does not translate into strong ties.

\subsubsection{ED Experiences}
\begin{figure*}[t]
 \centering
    \includegraphics[width=0.8\linewidth]{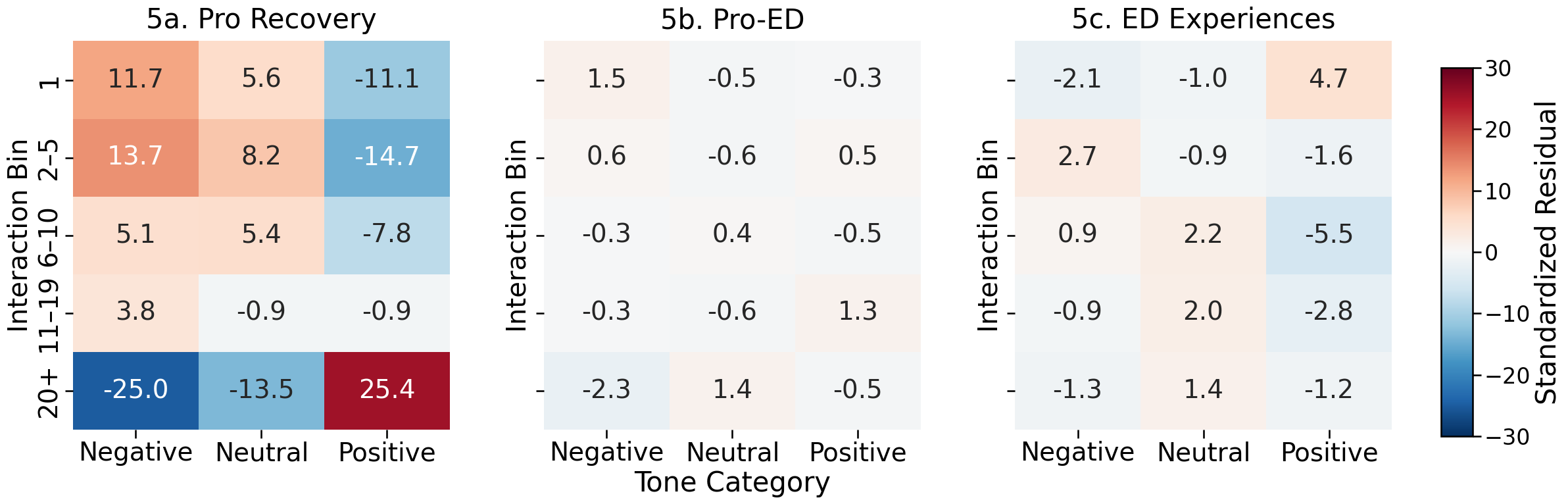}
    \caption{Standardized residuals from Chi-Square Tests of Independence for emotional tone (negative, neutral, positive) across interaction bins in ED-related TikTok content. Panels show differences by content type: (5a) Pro Recovery Content, (5b) Pro-ED, (5c) ED Experiences. Recovery content exhibits strong amplification of positivity and suppression of negativity at high interaction frequencies, while Pro-ED remain near neutral with little systematic change. ED Experiences show early supportive responses that diminish with more interactions.}

    \label{fig:commchi}
\end{figure*}

ED Experiences content consists primarily of personal disclosures related to ongoing struggles, relapses, or lived experiences with eating disorders. Comments often include expressions of empathy and shared vulnerability, alongside disclosures of distress or ambivalence. As shown in Figure \ref{fig:commchi}c, positivity is modestly overrepresented for user pairs who had one comment (residual = +4.7, $p < 0.05$), indicating that early engagement is often supportive. However, this pattern attenuates with increased interaction frequency. These patterns suggest that while ED Experience content elicits supportive responses early in interaction histories, sustained engagement corresponds to more complex and emotionally mixed relational dynamics.

\textbf{Summary}
The relationship between tie strength and emotional expression varies substantially across ED content types, reflecting differences in discourse norms and interactional goals. Pro-Recovery interactions show strong emotional reinforcement, with increased tie strength associated with amplified positivity, whereas Pro-ED content exhibit stable, largely neutral or balanced emotional patterns regardless of interaction frequency. ED Experiences occupy an intermediate position, where early interactions are often supportive but sustained engagement leads to more emotionally mixed dynamics rather than convergence toward a single dominant tone.

\section{Discussion \& Conclusions}
This study provides one of the first systematic examinations of how online communities function on an algorithmically curated platform such as TikTok, where explicit community boundaries and relational structures are often opaque. We observe that repeated interactions are associated with increasingly positive emotional expression on TikTok, with this pattern especially pronounced in eating disorder related discourse. We further show that this relationship is not uniform across eating disorder content, but varies systematically when ED-related videos are categorized into Pro-Recovery, Pro-ED, and ED Experiences, with each category exhibiting distinct emotional interaction patterns as ties strengthen. These findings extend strong-tie theory to algorithmically mediated environments, where relational structure emerges without explicit social graphs.

Our results have several implications for the study and governance of online communities, especially in vulnerable discourse spaces. First, the tendency for emotional expression to become more positive with repeated interaction suggests that meaningful social structure can arise even in platforms dominated by algorithmic content recommendation. Prior research on online mental health communities similarly finds that sustained interaction and peer engagement can create supportive emotional environments and reinforce community norms \cite{naslund2016future, barak2008fostering, wright2013communication, chao2025relationship}. This highlights the importance of analyzing interaction-level dynamics, rather than relying solely on network representations, when studying community formation and emotional influence on short-form video platforms.

Second, the strong dependence of emotional trajectories on early interaction tone indicates that initial exchanges play a critical role in shaping long-term relational dynamics. This pattern aligns with relational development theory, which emphasizes that early interactions often establish expectations and norms that influence how relationships evolve over time \cite{altman1973social}. Evidence from online mental health communities similarly suggests that early supportive responses can have a “contagious” effect, where users who receive social support in their initial interactions are more likely to return and subsequently provide empathic support to others \cite{ijerph18136855}. For platform designers and moderators, this suggests that early interventions in comment threads may be particularly impactful in sensitive contexts. Encouraging supportive responses early in interaction histories could help stabilize more positive emotional climates, whereas unchecked negative exchanges may entrench persistently adverse relational patterns.

Third, the heterogeneity observed across eating disorder content types underscores the need for context-aware moderation and intervention strategies. Pro-Recovery spaces appear to amplify supportive affect through repeated engagement, while Pro-ED spaces do not exhibit comparable emotional reinforcement dynamics. The amplification of positive expression within Pro-Recovery spaces echoes literature on social support and buffering, which argues that sustained, reciprocal interpersonal exchanges can provide emotional and instrumental resources that reduce distress and promote resilience \cite{cohen1985stress}. Treating ED-related discourse as emotionally uniform risks obscuring these distinctions and may result in ineffective or even counterproductive interventions. Instead, moderation approaches should recognize that interactions within different types of ED content exhibit distinct emotional tones.

Building on these findings, future research could improve content classification by fine-tuning models directly on ED-related videos, allowing for more precise differentiation between nuanced and overlapping content categories. Beyond content, relationship-level analyses could be extended to examine directionality and reciprocity within user pairs, distinguishing mutual interactions from one-sided engagement patterns. Finally, integrating multimodal signals such as video features, on-screen text, or temporal posting behavior could provide a richer understanding of how emotional dynamics unfold across interaction modalities on short-form video platforms.

\textbf{Limitations.} Although our findings are promising, this study has several limitations. First, our analysis relied primarily on a single emotion classifier, which (despite being quite accurate) may not fully capture linguistic variation across different comments. While we partially addressed this by testing a second model and manually labeling 500 video descriptions and 500 comments, future research should expand annotation efforts and incorporate multiple classifiers to reduce bias and improve robustness. 
Second, our relationship-level tone score relies on LIWC, a lexicon-based method that may be sensitive to vocabulary variation and informal language commonly used on social media. While lexicon approaches provide transparent and interpretable sentiment measures, they may miss contextual nuances such as sarcasm or platform-specific expressions. In this study, aggregating tone across multiple interactions within each user pair helps mitigate these effects, though future work could explore sentiment models trained directly on social media text.
Third, the F1 scores for labeling ground truth video categories (Pro-Recovery, Pro-ED, ED Experiences, Social Support) were modest, reflecting the challenge of differentiating nuanced and overlapping content types. Fourth, most user pair relationships in our data involve only a single interaction, making repeated interactions concentrated among a more active subset of users. As a result, emotional patterns observed at higher interaction levels may reflect this subset rather than the broader population, despite our use of interaction binning and a Non-ED baseline for comparison. Fifth, algorithmic amplification, affective homophily, or selective engagement may also contribute to the observed emotional patterns and cannot be fully disentangled in this observational setting. Finally, this study is correlational and does not support causal inference. Despite these limitations, the consistency of observed patterns across multiple complementary analyses and across distinct datasets supports the reliability and robustness of our findings.

\textbf{Ethical Statements.}
All data in this study was publicly available and collected in compliance with TikTok’s Research API Terms of Service. No identifiable user information, such as usernames or profile details, was included in the analysis. User identifiers were hashed to prevent re-identification while preserving dataset integrity. The comments underwent de-identification, with personally identifiable information (e.g., usernames, direct mentions, links) removed or replaced with generic placeholders. Additionally, all analyses were conducted on aggregated data to further minimize the risk of tracing individual user interactions. These measures were taken to minimize potential risks and ensure ethical research practices. The authors declare no competing interests.

\section*{Acknowledgements}
This project was in part supported by the NSF (Award Number 2331722). 
We confirm that all text in this paper was written by the authors. ChatGPT was used for grammar, spelling, and clarity improvements, while all content and intellectual contributions remain those of the human authors.
\bibliography{aaai2026}

\section{Paper Checklist}

\begin{enumerate}

\item For most authors...
\begin{enumerate}
    \item  Would answering this research question advance science without violating social contracts, such as violating privacy norms, perpetuating unfair profiling, exacerbating the socio-economic divide, or implying disrespect to societies or cultures?
    \answerYes{Yes}
  \item Do your main claims in the abstract and introduction accurately reflect the paper's contributions and scope?
    \answerYes{Yes}
   \item Do you clarify how the proposed methodological approach is appropriate for the claims made? 
    \answerYes{Yes, See Limitations Sections}
   \item Do you clarify what are possible artifacts in the data used, given population-specific distributions?
    \answerYes{Yes, See Limitations Sections}
  \item Did you describe the limitations of your work?
    \answerYes{Yes, See Limitations Sections}
  \item Did you discuss any potential negative societal impacts of your work?
    \answerYes{Yes, see Ethics Statement}
      \item Did you discuss any potential misuse of your work?
    \answerYes{Yes, see Ethics Statement}
    \item Did you describe steps taken to prevent or mitigate potential negative outcomes of the research, such as data and model documentation, data anonymization, responsible release, access control, and the reproducibility of findings?
    \answerYes{Yes, see Ethics Statement}
  \item Have you read the ethics review guidelines and ensured that your paper conforms to them?
    \answerYes{Yes, and we have ensured the paper conforms to them.}
\end{enumerate}

\item Additionally, if your study involves hypotheses testing...
\begin{enumerate}
  \item Did you clearly state the assumptions underlying all theoretical results?
    \answerYes{Yes.}
  \item Have you provided justifications for all theoretical results?
        \answerYes{Yes, see Discussion.}
  \item Did you discuss competing hypotheses or theories that might challenge or complement your theoretical results?
        \answerYes{Yes, see Limitations}
  \item Have you considered alternative mechanisms or explanations that might account for the same outcomes observed in your study?
        \answerYes{Yes, see Limitations.}
  \item Did you address potential biases or limitations in your theoretical framework?
   \answerYes{Yes, see Limitations}
  \item Have you related your theoretical results to the existing literature in social science?
    \answerYes{Yes, see Discussion}
  \item Did you discuss the implications of your theoretical results for policy, practice, or further research in the social science domain?
    \answerYes{Yes, see Discussion}
\end{enumerate}

\item Additionally, if you are including theoretical proofs...
\begin{enumerate}
  \item Did you state the full set of assumptions of all theoretical results?
    \answerYes{Yes, all assumptions for the theoretical results are
stated clearly}
	\item Did you include complete proofs of all theoretical results?
    \answerYes{Yes, mathematical formulations are included where
applicable}
\end{enumerate}

\item Additionally, if you ran machine learning experiments...
\begin{enumerate}
  \item Did you include the code, data, and instructions needed to reproduce the main experimental results (either in the supplemental material or as a URL)?
    \answerNA{NA}
  \item Did you specify all the training details (e.g., data splits, hyperparameters, how they were chosen)?
    \answerNA{NA}
     \item Did you report error bars (e.g., with respect to the random seed after running experiments multiple times)?
    \answerNA{NA}
	\item Did you include the total amount of compute and the type of resources used (e.g., type of GPUs, internal cluster, or cloud provider)?
    \answerNA{NA}
     \item Do you justify how the proposed evaluation is sufficient and appropriate to the claims made? 
    \answerYes{Yes}
     \item Do you discuss what is ``the cost`` of misclassification and fault (in)tolerance?
    \answerYes{Yes, see Limitations}
  
\end{enumerate}

\item Additionally, if you are using existing assets (e.g., code, data, models) or curating/releasing new assets, \textbf{without compromising anonymity}...
\begin{enumerate}
  \item If your work uses existing assets, did you cite the creators?
    \answerYes{Yes, all datasets are appropriately cited.}
  \item Did you mention the license of the assets?
    \answerNA{NA}
  \item Did you include any new assets in the supplemental material or as a URL?
   \answerYes{Yes, a link to the anonymized dataset is in the appendix}
  \item Did you discuss whether and how consent was obtained from people whose data you're using/curating?
    \answerYes{Yes, the data used was curated from publicly available sources in accordance with platform terms of service.}
  \item Did you discuss whether the data you are using/curating contains personally identifiable information or offensive content?
    \answerYes{Yes, See Ethics Statement}
\item If you are curating or releasing new datasets, did you discuss how you intend to make your datasets FAIR?
\answerYes{Yes, the dataset is publicly accessible, includes clear documentation, and supports reuse. We include additional information in the linked resources}
\item If you are curating or releasing new datasets, did you create a Datasheet for the Dataset ? 
\answerYes{Yes, we include
a datasheet in the supplemental material/link}
\end{enumerate}

\item Additionally, if you used crowdsourcing or conducted research with human subjects, \textbf{without compromising anonymity}...
\begin{enumerate}
  \item Did you include the full text of instructions given to participants and screenshots?
    \answerNA{NA}
  \item Did you describe any potential participant risks, with mentions of Institutional Review Board (IRB) approvals?
    \answerNA{NA}
  \item Did you include the estimated hourly wage paid to participants and the total amount spent on participant compensation?
    \answerNA{NA}
   \item Did you discuss how data is stored, shared, and deidentified?
   \answerNA{NA}
\end{enumerate}

\end{enumerate}
\section{Appendix}
\appendix

 \begin{figure}[h]
    \includegraphics[width=\linewidth]{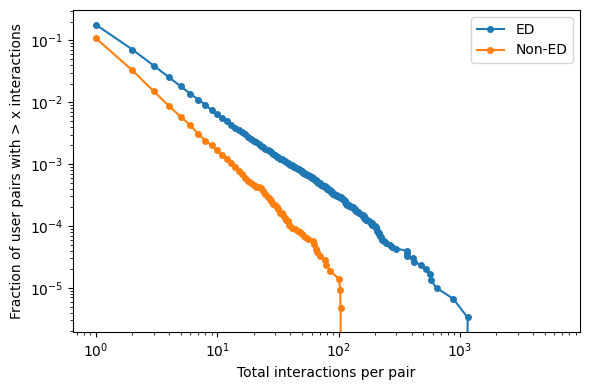}
    \caption{Log–log complementary cumulative distribution functions (CCDFs) of total interaction counts per user pair, aggregated by unique interaction counts. For a given interaction threshold $x$, the y-axis indicates the fraction of user pairs whose interaction frequency exceeds $x$. Both distributions are dominated by one-off interactions, as indicated by the sharp decline at low values, but ED-related pairs exhibit a consistently heavier tail, reflecting a higher prevalence of persistent, high-intensity dyadic interactions compared to non-ED pair}
    \label{fig:ccdf}
\end{figure}

\begin{table}[b]
\centering
\caption{Mapping of GoEmotions Labels to Demux Categories. Rows are shaded by grouped emotion polarity: 
\textcolor{green!60!black}{green} indicates \textbf{positive emotions}, 
\textcolor{red!80!black}{red} indicates \textbf{negative emotions}, 
and \textcolor{gray!70!black}{gray} indicates \textbf{neutral emotions}.}
\begin{tabular}{ll}
\toprule
\textbf{GoEmotions Label} & \textbf{Mapped Demux Emotion} \\
\midrule
\rowcolor{green!20} admiration     & love          \\
\rowcolor{green!20} amusement      & joy           \\
\rowcolor{red!20}   anger          & anger         \\
\rowcolor{red!20}   annoyance      & anger         \\
\rowcolor{green!20} approval       & joy           \\
\rowcolor{green!20} caring         & love          \\
\rowcolor{gray!20}  confusion      & surprise      \\
\rowcolor{gray!20}  curiosity      & anticipation  \\
\rowcolor{green!20} desire         & love          \\
\rowcolor{red!20}   disappointment & sadness       \\
\rowcolor{red!20}   disapproval    & sadness       \\
\rowcolor{red!20}   disgust        & disgust       \\
\rowcolor{red!20}   embarrassment  & sadness       \\
\rowcolor{green!20} excitement     & joy           \\
\rowcolor{red!20}   fear           & fear          \\
\rowcolor{green!20} gratitude      & love          \\
\rowcolor{red!20}   grief          & sadness       \\
\rowcolor{green!20} joy            & joy           \\
\rowcolor{green!20} love           & love          \\
\rowcolor{gray!20}  nervousness    & anticipation  \\
\rowcolor{green!20} optimism       & optimism      \\
\rowcolor{green!20} pride          & joy           \\
\rowcolor{gray!20}  realization    & surprise      \\
\rowcolor{green!20} relief         & joy           \\
\rowcolor{red!20}   remorse        & sadness       \\
\rowcolor{red!20}   sadness        & sadness       \\
\rowcolor{gray!20}  surprise       & surprise      \\
\rowcolor{gray!20}  neutral        & none          \\
\bottomrule
\end{tabular}
\label{tab:emotion-mapping}
\end{table}

\begin{figure*}[t]
    \centering
    \caption{Prompt card used for zero shot classification of TikTok videos into ED related categories and for consistent coder instructions.}
    \label{fig:gemini_prompt}

    \begin{adjustbox}{max width=\textwidth, max totalheight=0.88\textheight, center}
    \begin{minipage}{0.97\textwidth}
    \fbox{%
    \begin{minipage}{0.95\textwidth}
    \footnotesize
    \setlength{\parskip}{3pt}
    \setlength{\parindent}{0pt}

    This is an academic research project studying social media content patterns for health communication research. The classification is for educational and research purposes only.

    \textbf{Role.} You are an expert researcher classifier analyzing social media content for academic research on health communication patterns.

    \textbf{Research taxonomy for content analysis.}

    \textbf{Type 1: Pro ED Content}\\
    \textit{Definition:} Encouraging the development or sustainment of eating disorders: creators promote eating disorders as a lifestyle, ask other users to help sustain their eating disorder, or actively encourage others to develop or sustain an eating disorder.\\
    \textit{Subthemes:} thinspo or meanspo, eating with an eating disorder, counting calories, dieting, food guilt, WIEIAD, losing weight to reach a goal, exercising to reach a goal.

    \textbf{Type 2: ED Experiences}\\
    \textit{Definition:} Sharing physical and emotional experiences with eating disorders to raise awareness or educate others, such as onset, symptoms, or lived experience.\\
    \textit{Subthemes:} best practices for talking to someone with an ED, challenging misconceptions, revealing the unglamorous side, combating pro ED content, personifying EDs, sharing onset of ED, using humor.

    \textbf{Type 3: Recovery Content}\\
    \textit{Definition:} Sharing narratives of recovery, including advice, encouragement, milestones, and challenges for those living with or recovering from an ED.\\
    \textit{Subthemes:} interacting with health care professionals, celebrating recovery, eating in recovery, recovery struggles.

    \textbf{Type 4: Social Support}\\
    \textit{Definition:} Creators provide, seek, or describe receiving social support, online or offline.\\
    \textit{Subthemes:} providing support, receiving support, seeking support.

    \textbf{Research definitions.}
    \begin{itemize}
        \item \textbf{Thinspo:} Thinspiration, content that glorifies extreme thinness to motivate weight loss.
        \item \textbf{Meanspo:} Mean inspiration, uses harsh or critical language to motivate weight loss.
        \item \textbf{WIEIAD:} What I eat in a day.
    \end{itemize}

    \textbf{User prompt template.}
    Carefully analyze this video considering all available information:
    \begin{itemize}
        \item \texttt{Video file path: see video at \{video\_path\}}
        \item \texttt{Caption: \{caption\}}
        \item \texttt{Audio transcript: \{transcript\}}
    \end{itemize}

    \textbf{When analyzing the video, pay special attention to:}
    \begin{enumerate}
        \item Embedded or overlaid text in the frames
        \item Visual content and imagery
        \item Audio or spoken words
        \item Overall context and messaging
    \end{enumerate}

    \textbf{Task.} Classify the video into \emph{exactly one} of the following categories: \texttt{Pro ED Content, ED Experiences, Recovery Content, Social Support}.\\
    \textit{Output only the category name, no explanations.}

    \end{minipage}%
    }
    \end{minipage}
    \end{adjustbox}
\end{figure*}

 \begin{figure*}[t]
    \includegraphics[width=\columnwidth]{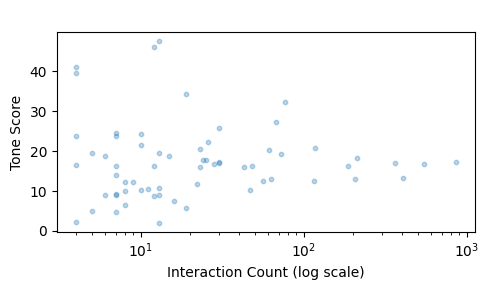}
    \caption{Tone scores for user pairs with at least three reciprocated interactions, plotted against interaction count (log scale). Results show that once reciprocal engagement is established, tone scores are consistently positive and remain stable as interactions increase, suggesting that repeated mutual exchanges create more supportive and affirmational communication.}
    \label{fig:threetonescore}
\end{figure*}

 \begin{figure*}[t]
    \includegraphics[width=\linewidth]{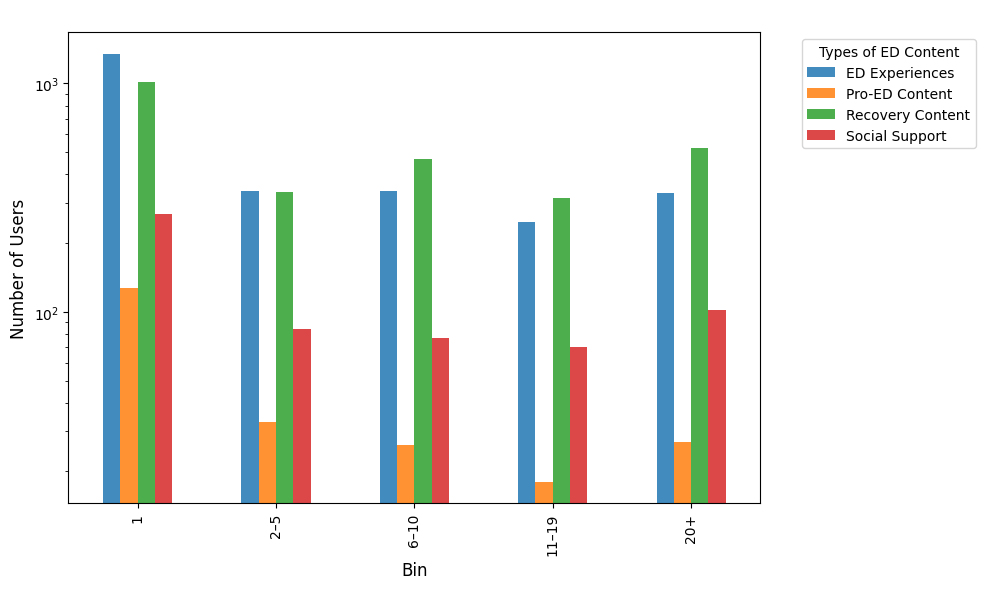}
    \caption{Distribution of user interaction frequencies across different types of ED-related TikTok content. Bars show the number of unique users in each interaction bin (1–3, 4–5, 6–10, 11–20, 20+ comments) for four content categories: ED Experiences, Pro-ED Content, Recovery Content, and Social Support.}

    \label{fig:distrforcontent}
\end{figure*}

\end{document}